\documentstyle[11pt,aaspp4]{article}

\def\scnot#1#2{#1 \times 10^{#2}}
\def\cm{{\rm cm}}

\def\kms{{\rm \;km\;s^{-1}}}
\def\kmsmpc{\kms\;{\rm Mpc}^{-1}}
\def\lya{Ly$\alpha$}

\def\taumax{\tau_{\rm max}}

\def\gam{\Gamma_{-12}}
\def\nh{n_{\sc HI}}
\def\nht{n_{\sc H}}
\def\nhtbar{\overline{n}_{\sc H}}
\def\be{\begin{equation}}
\def\ee{\end{equation}}
\def\Dbar{\overline{D}}
\def\Dmax{\overline{D}_{\rm max}}
\def\Bmin{B_{\rm min}}
\def\fhat{{\hat f}}
\def\that{{\hat \tau_c}}
\def\dtoh{({\rm D}/{\rm H})_P}
\def\tbar{\overline{T}_4}

\slugcomment{Submitted to ApJ}

\begin{document}

\title{A LOWER BOUND ON THE COSMIC BARYON DENSITY}
\author{
David H. Weinberg\altaffilmark{1}, 
Jordi Miralda-Escud\'e\altaffilmark{2}, 
Lars Hernquist\altaffilmark{3,4},
Neal Katz\altaffilmark{5}
}

\altaffiltext{1}{
Ohio State University, 
Department of Astronomy, Columbus, OH 43210; dhw@payne.mps.ohio-state.edu}
\altaffiltext{2}{
University of Pennsylvania,
Department of Physics and Astronomy, Philadelphia, PA 19104; 
jordi@dept.physics.upenn.edu}
\altaffiltext{3}{University of California, 
Lick Observatory, Santa Cruz, CA 95064; lars@ucolick.org}
\altaffiltext{4}{Presidential Faculty Fellow}
\altaffiltext{5}{Department of Astronomy, University of Washington, 
Seattle, WA 98195}

\begin{abstract}
We derive lower bounds on the cosmic baryon density from the requirement
that the high-redshift intergalactic medium (IGM) contain enough neutral
hydrogen to produce the observed \lya\ absorption in quasar spectra.
The key theoretical assumption that leads to these analytic bounds
is that absorbing structures are no more extended in redshift space than 
they are in real space.  This assumption might not hold if \lya\ clouds
are highly overdense and thermally broadened, but it is likely to hold
in the gravitational instability picture for the \lya\ forest suggested
by cosmological simulations, independently of the details of the
cosmological model.  The other ingredients that enter these bounds are an
estimate of (or lower limit to) the intensity of the photoionizing
UV background from quasars, a temperature $T \sim 10^4\;$K for the
``warm'' photoionized IGM that produces most of the \lya\ absorption,
a value of the Hubble constant, and observational estimates of the
mean \lya\ flux decrement $\Dbar$ or, for a more restrictive bound,
the distribution function $P(\tau)$ of \lya\ optical depths.
With plausible estimates of the quasar UV background and $\Dbar$,
the mean decrement bound implies a baryon density parameter
$\Omega_b \ga 0.0125 h^{-2}$, where $h \equiv H_0/100\;\kmsmpc$. 
With conservative values of the UV background intensity
and $\Dbar$, the bound weakens to $\Omega_b \ga 0.005 h^{-2}$, 
but the clustering of the absorbing gas that is 
required in order to reproduce the observed mean
decrement with this baryon fraction is incompatible with other
properties of quasar absorption spectra.  A recent observational
determination of $P(\tau)$ implies $\Omega_b \ga 0.0125 h^{-2}$
even for a conservative estimate of the quasar UV background,
and $\Omega_b \ga 0.018 h^{-2}$ for a more reasonable estimate.
These bounds are consistent with recent {\it low} estimates of the primordial
deuterium-to-hydrogen ratio $\dtoh$, which imply 
$\Omega_b \approx 0.025 h^{-2}$ when combined with standard
big bang nucleosynthesis.  Since the bounds account only for
baryons in the warm IGM, our results support earlier claims
that this component is the dominant reservoir of baryons in the
high-redshift universe.  The $P(\tau)$ bound on $\Omega_b$ is
incompatible with some recent high estimates of $\dtoh$
unless one drops the assumptions of standard big bang nucleosynthesis
or abandons the gravitational instability picture for the origin
of the \lya\ forest.
\end{abstract}


\section{Introduction}
Following the discovery of the first $z>2$ quasar (\cite{schmidt65}), 
Gunn \& Peterson (1965)
derived a stringent upper bound on the density of uniformly distributed,
neutral hydrogen in intergalactic space, by showing that the redshifted
\lya\ absorption of neutral gas with more than $\sim 10^{-4}$ of
closure density would turn quasar spectra virtually black at short
wavelengths, contrary to observation.  They concluded that the
intergalactic medium (IGM) must be highly ionized or extremely rarefied.
Within a few years, it became clear that the ubiquitous absorption lines
in quasar spectra are predominantly those of intervening neutral
hydrogen (\cite{lynds71}; \cite{sargent80}), and subsequent studies
have shown that these lines significantly depress the mean flux received from 
high-redshift quasars blue-ward of the \lya\ emission line
(\cite{oke82}; \cite{steidel87}; \cite{jenkins91};
Press, Rybicki, \& Schneider 1993, hereafter \cite{press93}; 
\cite{zuo93}; \cite{dobrzycki96}; \cite{rauch97}).
Furthermore, it is now recognized that the ambient ultraviolet (UV)
radiation background produced by high redshift quasars will strongly
photoionize gas near the cosmic mean density, so that a small amount
of diffuse neutral hydrogen corresponds to a much larger amount of
total hydrogen (e.g., Haardt \& Madau 1996,
hereafter \cite{haardt96}).  
In this paper, we will argue that matching the observed \lya\
absorption leads to interesting {\it lower} bounds on the mean baryon
density of the universe, which can be derived from quite general
assumptions about the state of the absorbing gas.

Recent cosmological simulations suggest that ``\lya\ forest'' lines arise
in diffuse, but non-uniform, intergalactic gas, and that they therefore
represent a phenomenon closely akin to the ``Gunn-Peterson effect''
(\cite{cen94}; \cite{petitjean95}; \cite{zhang95}; \cite{hkwm96};
for related semi-analytic modeling see \cite{bi93}; \cite{bi97}; \cite{hui97}).
Quantitative analyses show that these simulations require a high baryon
density in order to reproduce the observed mean opacity of the forest
(\cite{hkwm96}; \cite{miralda96}; \cite{cwkh97}; \cite{rauch97}; 
\cite{zhang97}).  For the UV background
predicted by HM based on the observed population of quasars, matching 
the mean opacity estimates of PRS typically requires $B \ga 2$, where
\be
B \equiv \frac{\Omega_b h^2}{0.0125} = \frac{\eta}{3.4 \times 10^{-10}}
\label{eqn:Bdef}
\ee
is the baryon density scaled to the fiducial big bang nucleosynthesis
estimate of Walker et al.\ (1991).  Here 
$h \equiv H_0/100\;\kmsmpc$, and $\eta$ is the baryon-to-photon ratio.

The bounds on $B$ derived in this paper will not be as high as
those derived from the simulations, but they have broader applicability
because they are not tied to a specific cosmological scenario, and
the simplicity of the arguments that lead to them makes it easier to
see how changes in the theoretical or observational inputs affect
the final result.  We do appeal to the simulations to motivate our one
crucial assumption, that structures with a volume filling factor $f$ in
real space have, on average, a filling factor no larger than $f$ in 
redshift space.  This assumption can also be phrased as a requirement
that typical \lya\ forest absorbers satisfy $X \geq 1$, where
\be
X \equiv \frac{{\rm real~space~extent}}{{\rm redshift~space~extent}} = 
\frac{H(z) d}{\Delta v}
\label{eqn:xdef}
\ee 
is the ratio of the Hubble flow across an absorber (with line-of-sight extent
$d$) to its line width $\Delta v$.
This assumption would not hold in a model of
spatially compact \lya\ clouds whose line widths are determined by
thermal broadening.  However, in the physical picture that emerges
from the simulations, the marginally saturated 
($N_{\rm HI} \sim 10^{13}-10^{15}\;\cm^{-2}$) absorption lines
that dominate the overall absorption usually arise in moderate overdensity
structures that are expanding in proper coordinates but contracting
in comoving coordinates (\cite{miralda96}, 1997).
The absorption line widths are determined largely by these coherent internal
motions rather than by thermal motions, and the velocity extent of
these features is generally no larger than the Hubble flow across them.  

Empirical support for the assumption that $X \geq 1$ comes from observations of 
quasar pairs, which yield a typical transverse coherence scale
$\sim 150\; h^{-1}\;{\rm kpc}$ for \lya\ forest systems at $z \sim 2$
(\cite{bechtold94p}; \cite{dinshaw94}, 1995).
For a typical line width $\Delta v \sim 25\;\kms$ (\cite{hu95}),
this lengthscale would imply $X \sim 3/R$ (for $\Omega=1$, $z=2$),
where $R$ is the ratio of transverse extent
to line-of-sight extent in real space.
If the absorbers are non-spherical, then
they are more often intercepted when they are closer
to ``face-on,'' so $R$ might reasonably exceed one on average.
However, the absorbers would have to be highly flattened,
coherent sheets in order to reproduce 
both the observed transverse coherence and the observed line widths
while having an average $X$ significantly smaller than one.
The structures giving rise to the \lya\ forest in the simulations
are commonly filamentary, with the transverse coherence scale
corresponding to the thickness of the filaments.

Rauch \& Haehnelt (1995) used the large transverse coherence scale
to argue that the \lya\ forest must contain a substantial fraction
of all baryons in the universe at high redshift.  Our arguments here are 
different from those of Rauch \& Haehnelt -- in particular, we work
directly from observed optical depths instead of from a derived HI
column density distribution --- but the spirit is similar.
In \S 2, we consider the lower bound on $B$ that can be obtained from
the mean \lya\ flux decrement (\cite{oke82}) alone.
In \S 3 we derive a more restrictive lower bound from the 
{\it distribution} of flux decrements (or equivalent optical depths),
recently measured from a set of seven Keck HIRES spectra by
Rauch et al.\ (1997).  Our analytic approach complements the direct
comparison to simulations carried out by Rauch et al.\ (1997), which
leads to stronger but less general bounds on the baryon density.
We discuss implications of our results in \S 4.

\section{A lower bound from the mean flux decrement}
A uniform IGM with neutral hydrogen density $\nh$ produces a \lya\ optical
depth
\be
\tau_u = \frac{\pi e^2}{m_e c} f_\alpha\lambda_\alpha H^{-1}(z) \nh,
\label{eqn:gp}
\ee
where $f_\alpha=0.416$ is the \lya\ oscillator strength and
$\lambda_\alpha=1216\AA$ is the transition wavelength
(\cite{gunn65}).
The Hubble parameter at redshift $z$ is
\be
H(z) = H_0 \left[\Omega_0 (1+z)^3 + (1-\Omega_0 -\lambda_0)(1+z)^2 + \lambda_0  
\right]^{1/2},
\label{eqn:hubble}
\ee
where $\lambda_0$ is the cosmological constant $\Lambda$ divided by
$3H_0^2$.
For realistic assumptions about the UV background, the IGM is highly
photoionized, and the neutral hydrogen density is
\be
\nh = \frac{\nht n_e \alpha(T)}{\Gamma} = \frac{1.16\nht^2\alpha(T)}{\Gamma},
\label{eqn:nh}
\ee
where $\alpha(T)$ is the recombination coefficient at the gas temperature $T$, 
$\Gamma$ is the photoionization
rate, and $\nht$ is the total hydrogen density.
Condition~(\ref{eqn:nh}) enforces balance between destruction of HI
by photoionization and creation by recombination.  In gas with 
$T \ga 10^5\;$K, collisional ionization enhances the destruction
rate and lowers $\nh$.
The mean value of $\nht$ is
\be
\nhtbar = \scnot{1.07}{-7} (1+z)^3 B\; \cm^{-3},
\label{eqn:nht}
\ee
with $B$ as defined in equation~(\ref{eqn:Bdef}).
Equations~(\ref{eqn:nh}) and~(\ref{eqn:nht}) assume a hydrogen mass
fraction $X=0.76$ and a helium mass fraction $Y=0.24$.
For gas at temperature $T_4 \equiv T/(10^4\;{\rm K}) \approx 1$,
the recombination coefficient is 
\be
\alpha(T) = \scnot{4.2}{-13} T_4^{-0.7} \cm^3 {\rm s}^{-1},
\label{eqn:alpha}
\ee
(\cite{abel97}).  Combining equations~(\ref{eqn:gp})--(\ref{eqn:alpha})
yields
\be
\tau_u = \scnot{2.31}{-4} (1+z)^5 (1+\Omega_0 z)^{-1/2} h^{-1} T_4^{-0.7}
\gam^{-1} B^2,
\label{eqn:tau_u}
\ee
where $\gam \equiv \Gamma/(10^{-12}\;{\rm s}^{-1})$ and we have
assumed $\Lambda=0$ to compute $H(z)$.
Equation~(\ref{eqn:tau_u}) agrees with, e.g., equation (36) of HM.

The mean \lya\ flux decrement produced by this uniform medium is
$\Dbar\equiv \langle 1-e^{-\tau} \rangle = 1-e^{-\tau_u} \equiv D_u$.
If the medium is optically thin (i.e., $\tau_u \ll 1$),
then clumping the gas tends to
increase $\Dbar$ because the mean neutral fraction at fixed temperature
increases in proportion to $\langle \nht^2 \rangle / \langle \nht \rangle^2$.
However, once the gas clumps produce absorption lines with optical
depths $\ga 1$, then $\Dbar$ is {\it decreased} by further clumping,
because more neutral atoms are added to saturated regions
(where they cannot contribute to increasing $\Dbar$),
at the expense of the interclump medium, where the resulting absorption
must decrease as matter is moved into the clumps.

Consider an idealized case in which all the gas collects into non-overlapping,
uniform density clumps, not necessarily spherical, which have a volume
filling factor $f$ and, thus, an overdensity $1/f$.  If a clump has the same
extent in redshift space as it does in real space, then it produces
an absorption line with optical depth $\tau_c=\tau_u/f^2$ and flux
decrement $D_c=1-e^{-\tau_c}$.  Since these lines fill a fraction $f$
of the spectrum, the mean flux decrement is
\be
\Dbar = fD_c = f\left(1-e^{-\tau_u/f^2}\right).
\label{eqn:Dbar}
\ee
In the optically thin limit, $\tau_u/f^2 \ll 1$, a Taylor expansion of
equation~(\ref{eqn:Dbar}) yields $\Dbar = D_u/f$.  In the saturated
line limit, $\tau_u/f^2 \gg 1$, equation~(\ref{eqn:Dbar}) gives $\Dbar=f$.
Thus, if $\tau_u \ll 1$, clumping the gas (lowering $f$) 
increases the absorption initially
but decreases it once lines become saturated, as expected from our
argument above.

It is straightforward to show (by setting $d\Dbar/df=0$) that the maximum 
value of $\Dbar$ in equation~(\ref{eqn:Dbar}) occurs when the optical
depth of the clumps is $\tau_u/\fhat^2 = \that = 1.25643...$;
$\that$ is the solution to the equation $\that=\frac{1}{2}(e^\that-1).$
If $\tau_u > \that$, then clumping can only decrease the overall absorption,
and $D_u$ is the maximum value of the mean flux decrement.
If $\tau_u < \that$, then the maximum decrement occurs for filling factor
\be
\fhat = \left(\tau_u/\that\right)^{1/2} = 0.89214\tau_u^{1/2},
\label{eqn:fhat}
\ee
implying a maximum decrement
\be
\Dmax = \fhat\left(1-e^{-\that}\right)= 0.71533\fhat= 0.63818\tau_u^{1/2}.
\label{eqn:Dmax}
\ee
This maximum decrement applies to any model with an arbitrary
distribution function of the gas density, because the general
case can be treated as a superposition of cases where all the gas
is in regions of constant density.
Thus, {\it the average absorption 
cannot exceed that of this optimal uniform clump case.}

For an observed value of $\Dbar$, equations~(\ref{eqn:tau_u}) 
and~(\ref{eqn:Dmax}) can be combined to give a lower bound on $B$:
\begin{eqnarray}
\Bmin = 
   &65.8 \left({\Dbar \over 0.63818}\right) 
    (1+z)^{-5/2}(1+\Omega_0 z)^{1/4} h^{1/2} T_4^{0.35} \gam^{1/2}, 
   &{\rm for}~\Dbar < 0.71533, \label{eqn:bound} \\
   &65.8\left[-{\rm ln}(1-\Dbar)\right]^{1/2} (1+z)^{-5/2}
    (1+\Omega_0 z)^{1/4} h^{1/2} T_4^{0.35} \gam^{1/2}, 
   &{\rm for}~\Dbar>0.71533.  \nonumber
\end{eqnarray}
The latter equation applies when $\Dbar > 1-e^{-\that}$, so that
the uniform IGM is the optimal case.  
Equation~(\ref{eqn:bound}) 
gives a lower bound on the baryon density 
on the assumption that
absorbing regions have, on average, the same filling factor
in redshift space as in real space.

If one assumes more generally that the absorbing regions have a constant
ratio $X$ of real space extent to redshift space extent,
then the optical depth of the clumps changes to $\tau_c=\tau_u X/f^2$
and the spectral filling factor to $f/X$ (or to unity if $f>X$).
One can carry through the same reasoning as above to find
that the bounds in equation~(\ref{eqn:bound}) 
simply change by the factor $X^{1/2}$.
As discussed in \S 1, cosmological simulations and quasar pair observations
suggest that $X \ga 1$ for systems that dominate the mean absorption,
making equation~(\ref{eqn:bound}) conservative.
Models with thermally broadened or expanding clouds could have $X<1$
and thereby escape the bound~(\ref{eqn:bound}), but the absorbers must
be inflated by a factor of four in redshift space in order to weaken
the bound by a factor of two.

In order to obtain numerical values of $\Bmin$, we must adopt values
of the mean flux decrement $\Dbar$,
the photoionization rate $\gam$, the gas temperature $T_4$, and
the cosmological parameters $\Omega_0$ and $h$.  
PRS, using the high-redshift quasar data of Schneider, Schmidt, \& Gunn (1994),
find $\Dbar(z)=1-e^{-\tau_{\rm eff}(z)}$ with
$\tau_{\rm eff}(z)=0.0037(1+z)^{3.46}$.  The absorption data used to 
derive this fit cover the redshift range $2.5 < z < 4.2$.
When extrapolated to $z=2$, this formula yields $\Dbar=0.15$, in agreement
with the value derived by Rauch et al.\ (1997) from Keck HIRES spectra.
Based on emission from
the observed population of quasars and reprocessing by the \lya\ forest,
HM find $\gam=1.4-1.6$ for $2<z<3$, and this value agrees well with
recent estimates from the \lya\ forest proximity effect 
(\cite{giallongo96}).  
In cosmological simulations with a photoionizing background,
Katz, Weinberg, \& Hernquist (1996) find $T_4 \approx 0.6$ for gas
at the cosmic mean density.
The solid line in Figure~1 shows the lower bound $\Bmin$ using
the PRS formula for $\Dbar(z)$, $\gam=1.4$, $T_4=0.6$,
$h=0.65$, and $\Omega_0=0.3$.  The derived constraint, $B \ga 1$,
is only weakly dependent on redshift in the range $2<z<3$ because the
observed dependence of $\Dbar$ on $z$ is canceled by the
redshift factors in equation~(\ref{eqn:bound}).
Our choice of temperature is conservatively low, since overdense gas
is usually hotter than gas at the mean density, and since
the Katz et al.\ (1996) simulations do not incorporate
heat injection during reionization (\cite{miralda94}).  
Including reionization heating, whose magnitude
is theoretically uncertain, could raise the gas temperatures 
enough to increase $\Bmin$ by $20-50\%$.

\begin{figure}
\centerline{
\epsfxsize=4.0truein
\epsfbox[90 415 460 730]{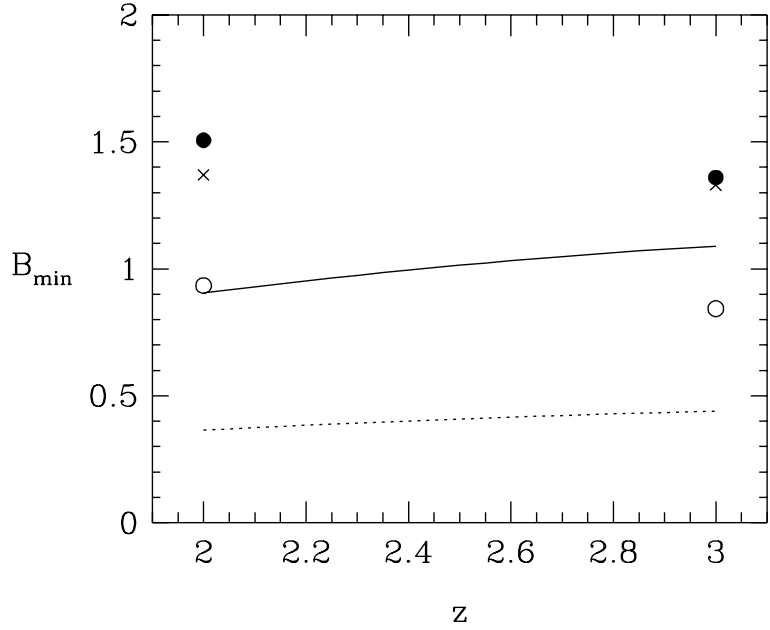}
}
\caption{
Lower bounds on the scaled baryon density $B\equiv \Omega_b h^2/0.0125$,
obtained from the mean flux decrement (solid and dotted lines,
eq.~[\ref{eqn:bound}]) and from the optical depth distribution
(filled and open circles, eq.~[\ref{eqn:bound2}]).
The solid line shows the mean decrement bound with
the PRS formula for $\Dbar(z)$, $\Omega_0=0.3$, $h=0.65$, 
$T_4=0.6$, and $\gam=1.4$. 
The dotted line shows the mean decrement bound for conservative
parameter choices: $\Dbar(z)$ equal to 65\% of the PRS values,
$h=0.5$ and $\gam=0.7$.
Filled circles show the $P(\tau)$ bounds at $z=2$ and $z=3$, using 
$\beta=0.633$, $\Omega_0=0.3$, $h=0.65$, $\tbar  =0.6$, and $\gam=1.4$.
Open circles show the corresponding bounds for $h=0.5$ and $\gam=0.7$.
Crosses show Rauch et al.'s (1997)
estimates of $B$ for $h=0.5$ and $\gam=0.7$,
obtained by comparing the observed optical
depth distribution to that derived from a numerical simulation of
the standard cold dark matter model.
}
\end{figure}

Of the various inputs to the lower bound in equation~(\ref{eqn:bound}),
$\Bmin$ is most sensitive to the mean flux decrement $\Dbar$ itself.
As already mentioned, the value obtained by Rauch et al.\ (1997)
at $z=2$ agrees almost perfectly with the PRS formula that we have
used.  At $z=3$ the PRS formula predicts $\Dbar=0.36$, and
Rauch et al.\ (1997) measure $\Dbar=0.32$, which rises to $\Dbar=0.35$
after including a theoretically estimated correction for continuum 
fitting bias.  Steidel \& Sargent (1987) find $\Dbar=0.24$ for a 
sample of seven quasars with a mean absorption redshift of 2.64;
the PRS formula yields $\Dbar=0.28$ at this redshift.
These three studies thus agree to 15\% or better.
However, Zuo \& Lu (1993) and Dobrzycki \& Bechtold (1996) find
values of $\Dbar$ that are typically about 35\% smaller than the PRS values.
The Rauch et al.\ (1997) values seem the most secure because they
are based on the highest quality quasar spectra, but the sample used
is relatively small, so further analyses of spectra with similar
resolution and signal-to-noise ratio will be needed to arrive at
a definitive measurement.

The other significant uncertainty in $\Bmin$ is the value of $\gam$.
The HM estimate is more likely to be too low than too high, since
it assumes that quasars are the only source of the UV background.
However, HM do include an extrapolation of the quasar luminosity 
function to allow for faint sources below existing survey limits,
and Rauch et al.\ (1997) conclude that $\gam$ could be up to a factor
of two below HM's estimate, at the price of worsening the agreement
with proximity effect estimates (e.g., \cite{bajtlik88};
\cite{bechtold94}; \cite{giallongo96}; \cite{cooke97}).
The dotted line in Figure~1 shows the value of $\Bmin$ obtained
after reducing $\Dbar(z)$ to 65\% of the PRS values, lowering
$\gam$ to 0.7, and, for good measure, dropping $h$ to 0.5. 
With all of the parameters in equation~(\ref{eqn:bound}) pushed to
these favorable values, the lower bound is $\Bmin \approx 0.45$.
Of course, achieving the observed absorption with $B=\Bmin$ 
requires that all of the baryons be collected into uniform density,
6000 degree gas clumps that have $\tau=\that$, which further
requires that the overdensity of these clumps evolve with redshift as
$1/f = 0.71533/\Dbar(z)$ (see equation~[\ref{eqn:Dmax}]).

\section{A lower bound from the optical depth distribution}
The clustering pattern required to achieve the 
bound of equation~(\ref{eqn:bound}) is not just physically contrived,
it is inconsistent with the observed properties of quasar spectra.
These show absorption at a range of optical depths, not a
set of $\tau=1.25$ lines separated by absorption-free regions.
The two cosmological simulations discussed in Rauch et al.\ (1997)
produce spectra whose optical depth distributions agree well with
observations, and Rauch et al.\ argue that they therefore provide reliable
models of the IGM with which to derive $B$ given values of $\gam$ and $h$.
Here we extend the argument of \S 2 to derive a lower bound on $B$
directly from the observed optical depth distribution, based on
some simple assumptions about the IGM.

If we ignore peculiar velocities and thermal broadening, 
then the optical depth of gas with
overdensity $(\nht/\nhtbar)$ and temperature $T$ is
\be
\tau = \tau_u \left(\frac{\nht}{\nhtbar}\right)^2 
       \frac{\alpha(T)}{\alpha(T_u)},
\label{eqn:tau}
\ee
where $\tau_u$ and $T_u$ are the optical depth and temperature that
the gas would have if it were uniformly distributed.  The recombination
coefficient $\alpha(T) \propto T^{-0.7}$ for the temperature range
of interest.  If we assume a relation between density and temperature,
then we can associate a density with each optical depth $\tau$.
In cosmological simulations that 
adopt $B=1$ and the HM background spectrum, Weinberg, Hernquist, \& Katz (1997)
find $T_4 \approx 0.6 (\rho_b/\bar\rho_b)^{0.6}$ for the gas that dominates
the absorption, with higher temperatures in collapsed, shock heated regions.
The $T \propto \rho^{0.6}$ relation arises because denser gas absorbs 
energy from the photoionizing background more rapidly
(see further discussion by \cite{cwkh97}; \cite{hg97}; \cite{mwhk97}).
Reionization heating (\cite{miralda94})
can raise the temperature of the gas independent
of density and thus tends to produce a higher multiplicative constant
and a weaker trend with density.
We will assume
\be
T_4 = \tbar   (\nht/\nhtbar)^{\gamma},
\label{eqn:trho}
\ee
which implies
\be
\nht = \nhtbar (\tau/\tau_u)^\beta,\quad {\rm where}~ 
       \beta\equiv (2-0.7\gamma)^{-1}.
\label{eqn:density}
\ee
Theoretically plausible ranges at $z=2-3$ are $T_4 \sim 0.5-1.5$ and 
$\gamma \sim 0.0 - 0.6$ ($\beta \sim 0.5-0.65$; see \cite{hg97}).
Since shocked gas is heated to higher temperatures
and produces less absorption, assuming that all of the gas
lies on the temperature-density relation tends to underestimate
the amount of baryons present.

Let $P(\tau)$ denote the optical depth probability distribution, so that
$P(\tau)d\tau$ is the probability that a randomly selected point on 
a spectrum has optical depth in the range $\tau \rightarrow \tau +d\tau$.
We can use equation~(\ref{eqn:density}) to 
compute the mean density by integrating over $P(\tau)$:
\be
\nhtbar = \int_0^\infty \nht(\tau) P(\tau) d\tau = 
\nhtbar \tau_u^{-\beta} \int_0^\infty \tau^\beta P(\tau) d\tau,
\ee
and thus
\be
\tau_u = \left[\int_0^\infty \tau^\beta P(\tau) d\tau\right]^{1/\beta},
\label{eqn:taubeta}
\ee
where $\tau_u$ depends on $B$ and other parameters as indicated in 
equation~(\ref{eqn:tau_u}).
We have relied on our assumption of Hubble flow broadening both
to associate a physical density with a spectral optical depth and to identify a
probability $P(\tau)d\tau$ in redshift space with an equal filling
factor in real space.  We can crudely incorporate a more general
model by assuming that the ratio $X$ of real space extent to redshift
space extent is a function of $\tau$ alone, $X=X(\tau)$.
If $X(\tau)>1$, then ``squeezing'' in redshift space has enhanced the
optical depth by a factor $X(\tau)$, but the spectral filling factor
$P(\tau)d\tau$ corresponds to a larger real space filling factor
$X(\tau)P(\tau)d\tau$.  Thus,
\be
\nhtbar = \int_0^\infty \nht(\tau) P(\tau) d\tau = 
\nhtbar \tau_u^{-\beta} \int_0^\infty [\tau/X(\tau)]^\beta X(\tau)P(\tau) d\tau,
\ee
implying
\be
\tau_u = \left[\int_0^\infty [X(\tau)]^{1-\beta} 
	 \tau^\beta P(\tau) d\tau\right]^{1/\beta}.
\label{eqn:xtaubeta}
\ee
It is conceptually helpful to rewrite this equation in the form
\be
\tau_u = X_w^{(1-\beta)/\beta} \left[\int_0^\infty
	 \tau^\beta P(\tau) d\tau\right]^{1/\beta},
\label{eqn:xtaubeta2}
\ee
where $X_w$ is the value of $X$ weighted by its contribution to
the integral~(\ref{eqn:xtaubeta}):
\be
X_w \equiv \left[
    \frac{\int_0^\infty [X(\tau)]^{1-\beta}\tau^\beta P(\tau) d\tau}
         {\int_0^\infty \tau^\beta P(\tau) d\tau}
	 \right]^{1/(1-\beta)}.
\label{eqn:xwdef}
\ee
In the picture suggested by cosmological simulations, $X(\tau)$
is typically smaller than unity for small $\tau$, since
the lowest optical depths arise primarily in underdense regions
that are expanding faster than the Hubble flow.  It then rises above
unity for intermediate $\tau$, where the gas is expanding slower than
Hubble flow as it falls into overdense structures.
Saturated absorption lines are often thermally broadened even in
the cosmological simulations, so $X(\tau)$ may fall below unity again
at high $\tau$.  Averaging over these three regimes, $X_w \approx 1$
is a fair approximation for gravitational instability
models of the \lya\ forest (see below).  Values of $X_w \ll 1$ would require
dense clouds that were thermally broadened or expanding rapidly
relative to Hubble flow.

We can combine equation~(\ref{eqn:xtaubeta2}) with equation~(\ref{eqn:tau_u})
for the uniform optical depth to obtain a new lower bound on $B$,
\be
\Bmin = 65.8\; X_w^{(1-\beta)/2\beta}\;
\left[\int_0^\infty \tau^\beta P(\tau) d\tau\right]^{1/2\beta}
(1+z)^{-5/2} (1+\Omega_0 z)^{1/4} h^{1/2} \tbar  ^{0.35} \gam^{1/2}. 
\label{eqn:bound2}
\ee
If all baryons were in the form of intergalactic gas that followed
the temperature-density relation~(\ref{eqn:trho}), then 
equation~(\ref{eqn:bound2}) would be an estimate of the
baryon density rather than a lower bound.  
However, some fraction of the baryons should be in stars, 
shock heated gas, and (possibly) baryonic dark matter.
Furthermore, since it is difficult to measure the optical depth
once $e^{-\tau}$ is close to zero,
we will in practice have to compute the integral
in~(\ref{eqn:bound2}) with a conservative assumption, e.g., that 
all regions with $\tau > \taumax \sim 3$ 
(transmission less than 0.05) have $\tau=\taumax$.
Equation~(\ref{eqn:bound2}) therefore leads to 
a lower bound on $B$ rather than an estimate.

Note that if we adopt the optimal uniform density clump model from \S 2,
then $P(\tau) = \fhat\delta(\that) + (1-\fhat)\delta(0)$, where $\delta(x)$
denotes the Dirac-delta function. 
With $\beta=1/2$, the integral in equation~(\ref{eqn:bound2})
becomes $\that^{1/2}\fhat$, which, by equation~(\ref{eqn:Dmax}),
is $\that^{1/2}\Dbar/(1-e^{-\that}) = (\Dbar/0.63818).$
With $X_w=1$, we recover the bound~(\ref{eqn:bound}) from \S 2,
as expected.  The maximum value of $(1-e^{-\tau})/\tau^{1/2}$ occurs
when $\tau=\that$, so for
any other optical depth distribution that has the
same mean decrement $\Dbar = \int_0^\infty(1-e^{-\tau})P(\tau)d\tau$, 
the factor $\int_0^\infty \tau^{1/2} P(\tau) d\tau$ must exceed 
$(\Dbar/0.63818)$,
and equation~(\ref{eqn:bound2}) yields a more restrictive lower bound on $B$
than equation~(\ref{eqn:bound}).
This derivation from the optical depth probability distribution 
is an alternative route to~(\ref{eqn:bound}).

We have checked that equation~(\ref{eqn:bound2}) with $X_w=1$
gives a lower bound to $B$ in realistic models of the \lya\ forest
by applying it to spectra from Croft et al.'s (1997) simulations
of three cold dark matter (CDM) cosmological models with $B=1$.
We measure $P(\tau)$ from the simulations at $z=2$ and $z=3$, 
in each case using the the value of $\gam$ required to match the PRS mean 
flux decrement.  [See figure~11 of \cite{cwkh97} for plots of
$P(\tau)$ at $z=2.33$.]
We then apply equation~(\ref{eqn:bound2}), assuming that 
all regions with $\tau>3$ have $\tau=3$ in order to account 
for the limited ability of realistic data to estimate optical depths 
in saturated regions.  
The derived lower bound is about 70\% of the models' true baryon density
in the ``standard'' (SCDM, $\Omega=1$, $h=0.5$, $\sigma_8=0.7$)
and open (OCDM, $\Omega_0=0.4$, $h=0.65$, $\sigma_8=0.75$) models
and about 50\% of the true baryon density in the COBE-normalized
$\Omega=1$ model (CCDM, $\Omega=1$, $h=0.5$, $\sigma_8=1.2$).
The bounds are insensitive to the value of $\tau_{\rm max}$, changing
by at most 5-10\% if $\tau_{\rm max}$ is changed to 2 or to 4.
The baryons ``missed'' by equation~(\ref{eqn:bound2}) are those
that have been shock heated to high temperatures, condensed into very
high density clumps with $\tau \gg 3$ (giving rise to Lyman limit
and damped \lya\ systems), or converted into stars.  
These are a larger fraction of the total baryon density in the CCDM
model because of its higher mass fluctuation amplitude.
For the SCDM model, we have estimated $X_w$ directly by comparing
the derived values of $\Bmin$ to values derived from spectra that include no 
thermal broadening or peculiar velocity distortions (and thus
have $X_w=1$ by construction).  We find $X_w=1.15$ at $z=2$ and 
$X_w=1.35$ at $z=3$, so while infall (in comoving coordinates) 
leads to $X_w > 1$ as expected, assuming $X_w=1$
only reduces $\Bmin$ by 4\% at $z=2$ and 9\% at $z=3$ in this model.

Rauch et al.\ (1997) have measured the cumulative distribution of
\lya\ flux decrements from Keck HIRES spectra of seven quasars,
whose emission redshifts range from 2.5 to 4.5.
They find that artificial spectra from the SCDM simulation mentioned
above reproduce the observed flux decrement distribution quite
accurately, once $\gam$ is chosen so that the mean decrement
matches that derived from the data.
The artificial (and real) spectra include noise and are measured with
a locally estimated continuum level.  
Since we would like to use the true optical depth distribution, free of 
observational artifacts, to obtain $\Bmin$, we measure $P(\tau)$
directly from the noiseless simulated spectra and use it as a 
surrogate for the observed $P(\tau)$ in equation~(\ref{eqn:bound2}).
As before, we set $\tau=3$ in the regions where $\tau>3$.
These high optical depth regions 
cover 9\% of the spectrum at $z=3$ and 2.5\% at $z=2$.

The filled circles in Figure~1 show the bound $\Bmin$ obtained from
equation~(\ref{eqn:bound2}) at $z=2$ and $z=3$ assuming $X_w=1$,
$\beta=0.633$ (i.e., $\gamma=0.6$), $\Omega_0=0.3$, 
$h=0.65$, $\tbar  =0.6$, and $\gam=1.4$.
These assumptions correspond to those used in obtaining the solid
line, and, as expected, the values of $\Bmin$ derived from the optical
depth distribution are higher than those derived from the mean
flux decrement alone.  At $z=3$ the bound increases by about 25\%
to $\Bmin=1.35$, and at $z=2$ it increases by nearly 70\% to $\Bmin=1.5$.
The open circles show the bounds derived for $h=0.5$ and $\gam=0.7$ 
(with other parameters unchanged):
$\Bmin=0.85$ at $z=3$ and $\Bmin=0.95$ at $z=2$.  These are substantially
stronger than the mean decrement bounds represented by the dotted line,
which also assume $h=0.5$ and $\gam=0.7$.
However, in going from the solid line to the dotted line we reduced
the mean flux decrements by 35\%, while the filled and open circles
are both based on the Rauch et al.\ (1997) measurements of the
optical depth distribution.
If these measurements are accurate (the small sample size being their
primary limitation),
then the mean baryon density cannot be much smaller than $B \sim 0.9$
unless $\gam$ is smaller than 0.7, $h$ is smaller than 0.5, or
\lya\ absorbers are substantially larger in redshift space than in
real space so that $X_w < 1$.  The crosses in Figure~1 show the
values of $B$ that Rauch et al.\ (1997) obtain by requiring that the
SCDM simulation match the observed $P(\tau)$ with $\gam=0.7$ and $h=0.5$.
These lie about a factor 1.5 above the open circles because the
direct comparison to the simulation allows these estimates to include
the contribution of stars and shock heated gas.

\section{Discussion}
The bound on the baryon density from the mean flux decrement,
equation~(\ref{eqn:bound}), and the bound from the optical depth
distribution, equation~(\ref{eqn:bound2}), are the principal results
of this paper.  The mean decrement bound leads to $B \ga 1$ if one
takes the most convincing estimates of $\Dbar$ and of the photoionization
rate $\gam$ from quasars.  With the Rauch et al. (1997) determination
of $P(\tau)$, the optical depth distribution bound gives $B \ga 1$ 
even if one reduces $\gam$ to half of the best estimate value.
These constraints imply a baryon-to-photon ratio $\eta \ga 3.4\times 10^{-10}$
and, assuming the standard model for big bang nucleosynthesis, a corresponding
primordial deuterium-to-hydrogen ratio $\dtoh \la 6 \times 10^{-5}$.
This constraint is consistent with the estimate 
$\dtoh = 2.3 \pm 0.4 \times 10^{-5}$ of Tytler, Fan, \& Burles (1996) from 
a high-redshift Lyman limit system, but it is inconsistent with the
estimates of $\dtoh \ga 10^{-4}$ obtained from other high-redshift Lyman
limit systems by Songaila et al.\ (1994), Carswell et al.\ (1994),
and Rugers \& Hogan (1996ab).
Current observational estimates of $P(\tau)$, $\gam$, and $h$ can only
be reconciled with the high $\dtoh$ estimates by abandoning the
gravitational instability picture of the \lya\ forest and returning
to a scenario of dense, thermally broadened 
clouds with $X_w < 1$, or, more radically, by abandoning standard
big bang nucleosynthesis.

Given an estimated value of $\gam$,
equation~(\ref{eqn:bound2}) provides an estimate of the number
of baryons in the warm ($T \sim 10^3 - 10^5\;$K), diffuse
($\rho/{\overline \rho} \la 10$) intergalactic medium.
With the Rauch et al.\ (1997) $P(\tau)$, HM's value $\gam=1.4$, 
and plausible choices for other parameters (see Fig.~1 caption),
one obtains $B_{\rm IGM} \approx 1.4$ 
at $z \sim 2-3$ (filled circles in Fig.~1).
Within standard big bang nucleosynthesis, one must generously increase
the estimated observational errors in the primordial $^4{\rm He}$ abundance 
even to accommodate $B$ as large as 2 (see, e.g., \cite{hata95}),
and measurements of $({\rm D}/{\rm H})$ in the local interstellar
medium (see, e.g., \cite{mccullough92}; \cite{linsky95}) imply $B \la 2.4$.
This high value of $B_{\rm IGM}$ therefore
suggests that the warm IGM contains most of the baryons
in the universe at these redshifts, as predicted by the cosmological
simulations and as argued by Rauch \& Haehnelt (1995) on different,
but physically related, empirical grounds.

While most of the high-redshift hydrogen resides in the warm,
photoionized IGM, most of the {\it neutral} hydrogen resides in
high column density, damped \lya\ systems.
Because we truncate $P(\tau)$ at $\tau_{\rm max}=3$, 
equation~(\ref{eqn:bound2}) 
severely undercounts the gas in these systems,
which contribute $B_{\rm DLA} \sim 0.1-0.3 h^2$ at $z \sim 3$ 
(\cite{wolfe95}).
Our lower limit to $B_{\rm IGM}$ far exceeds the baryon density
of stars in bright galaxies today, $B \sim 0.16 h^2$ (\cite{persic92}).  
Thus, the baryons that were in the warm IGM at $z=2$ must either 
(a) remain in the IGM today, (b) have formed brown dwarfs or 
another form of baryonic dark matter since $z=2$, or (c) have formed
stars in systems of very low surface brightness that have been
missed in standard estimates of the galaxy luminosity function.

There are a number of anticipated observational developments that might
strengthen (or weaken) one's confidence in the bounds plotted in Figure~1.
The most important will be determinations of $\Dbar$ and $P(\tau)$
from larger samples of high resolution, high signal-to-noise ratio
quasar spectra, since the values of $\Bmin$ depend primarily on these
observational inputs.  Ongoing quasar surveys and, in a few years,
the Sloan Digital Sky Survey, will yield improved determinations of
the quasar luminosity function, which can be combined with the HM
formalism to yield more definitive estimates of
the quasar contribution to the photoionizing background. 
Further analyses of quasar pairs and studies of absorption line shapes
in high resolution spectra may provide more compelling evidence
for extended \lya\ forest absorbers that are broadened largely by Hubble flow,
as predicted by cosmological simulations.
Such observations would support our key theoretical assumption that
$X_w \ga 1$, and they would reinforce the view that the \lya\
forest arises in a smoothly fluctuating intergalactic medium
that is the dominant reservoir of high redshift baryons.

\acknowledgments
We thank Michael Rauch and Rupert Croft for helpful discussions.
We acknowledge support from NASA Grants NAG5-3111 and NAG5-3525
and from the NSF through grant ASC 93-18185 and the Presidential
Faculty Fellows Program.  Computing support was provided by the
San Diego, Pittsburgh, and Illinois Supercomputing Centers.

\end{document}